\documentclass[a4paper,10pt,twoside]{paper}
\usepackage{multicol}
\usepackage{graphicx}
\usepackage{booktabs}
\usepackage{amssymb,bm,mathrsfs,bbm,amscd}
\usepackage[tbtags]{amsmath}
\usepackage{lastpage}
\usepackage{lineno}

\begin{document}

 \fancyhead[c]{\small Submitted to 'Chinese Physics C'} 

  \title{Simulation of natural radioactivity backgrounds in the JUNO central
  detector \thanks{This work is supported by the Strategic Priority Research Program of the Chinese Academy of Sciences, Grant No. XDA10010900; the CAS Center for Excellence in Particle Physics (CCEPP); National Natural Science Foundation of China (NSFC); the Chinese Academy of Sciences (CAS) Large-Scale Scientific Facility Program; Joint Large-Scale Scientific Facility Funds of the NSFC and CAS under Contracts Nos. U1332201.}}

    \author{Xinying Li$^{1,2}$\email{lixy@ihep.ac.cn}
      \quad Ziyan Deng$^{2}$
      \quad Liangjian Wen$^{4}$
      \quad Weidong Li$^{2}$
      \quad Zhengyun You$^{3}$
      \quad Chunxu Yu$^{1}$\\
      \quad Yumei Zhang$^{3}$
      \quad Tao Lin$^{2}$
    }
\maketitle

\address{
    $^1$ School of Physics, NanKai University, Tianjin 300071, China\\
    $^2$ Institute of High Energy Physics, Chinese Academy of Sciences, Beijing 100049, China\\
    $^3$ Sun Yat-sen University, Guangzhou 510275, China\\
    $^4$ State Key Laboratory of Particle Detection and Electronics(Institute of High Energy Physics, CAS)\\
        }

\begin{abstract}
   The Jiangmen Underground Neutrino Observatory (JUNO) is an experiment proposed to determine the neutrino mass hierarchy and probe the fundamental properties of neutrino oscillation.  The JUNO central detector is a spherical liquid scintillator detector with 20 kton fiducial mass. It is required to achieve a $3\%/\sqrt{E(MeV)}$ energy resolution with very low radioactive background, which is a big challenge to the detector design. In order to ensure the detector performance can meet the physics requirements, reliable detector simulation is necessary to provide useful information for detector design. A simulation study of natural radioactivity backgrounds in the JUNO central detector has been performed to guide the detector design and set requirements to the radiopurity of detector materials.

\end{abstract}




    \begin{multicols}{2}

    \section{Introduction}
   JUNO\cite{lab1} is a multipurpose neutrino experiment designed to determine the neutrino mass hierarchy and precisely measure the oscillation parameters by detecting reactor neutrinos from the Yangjiang and Taishan nuclear power plants. It also intends to observe supernova neutrinos, study the atmospheric, solar neutrinos and geo-neutrinos, and perform exotic searches.

    JUNO is located in Kaiping, Jiangmen, in Southern China. It is about 53 km away from Yangjiang and Taishan nuclear power plants, both of which are under construction. The planned total thermal power of these reactors is 36 GW. In addition, there are no other nuclear power plants within 200 km.  In order to suppress the backgrounds induced  by cosmic ray muons, the detector is deployed underground with a total overburden of 700 m rock. The current JUNO detector design consists of a liquid scintillator central detector, a water cherenkov detector and a muon tracker.

    In recent years, liquid-scintillator detectors have made  important contributions to low-energy
    neutrino physics\cite{lab2, lab3, lab4, lab5, lab6, lab7}. 
    Currently the JUNO central detector with 35.4 m diameter is the largest liquid scintillator detector.
    We use Geant4\cite{lab8} simulation to evaluate the design and optimize the dimensions of the detector. 
    Geant4 also shows a good ability to simulate the backgrounds induced by the materials.

    \section{Several options of central detector}
     The challenging construction of the JUNO central detector includes the inner transparent tank and the outer supporting structure.
     To get the energy resolution as good as $3\%/\sqrt{E(MeV)}$, the huge detector has to optimize the collection of optical photons from the liquid-scintillator (LS) target while suppressing the variety of background sources. 
    
     The collection of optical signals is mainly determined by the light yield and transparency of LS, and the coverage and quantum efficiency(QE) of the PMT photocathode.
     The PMTs photocathode coverage should be greater than 75\%.
     In the beginning, there were several design options for the detector: acrylic option, module option and balloon option. 
     In order to better compare the pros and cons of each option, PMTs are fixed on the same radius.
     The PMT coverage is approximately the same in these detector options.
     The main differences between these detector options lie in the arrangement method of the PMTs and the buffer material. 
     In these simulations, the light yield of LS is 10400 photons/MeV, and the QE of PMT is 35\%. 
     The attenuation length of LS is 20 m at the wavelength of 430 nm, which corresponds to an absorption length of 60 m with a rayleigh  scattering length of 30 m.
      Detailed explanations of these options are given in the following subsections.  
    
    \subsection{Acrylic option}
    In this option, the inner acrylic tank is spherically shaped to hold 20 kton LS.
    The acrylic sphere is supported by the stainless steel truss while the truss is held by some supporting legs at the bottom of the water pool in the experiment hall. To reduce the radioactivity background, oxygen-free copper is used for the joint between stainless steel truss and acrylic tank. The thickness of the acrylic tank is 12 cm.
    Between the acrylic sphere and the truss, PMTs are mounted inward facing to the truss to detect the optical signal from LS.
    Fig.~\ref{fig1} shows the acrylic tank and the steel truss of the central detector. Ultrapure water is filled in as shielding liquid outside of the acrylic tank.
   
    \begin{center}
    \includegraphics[width=6cm]{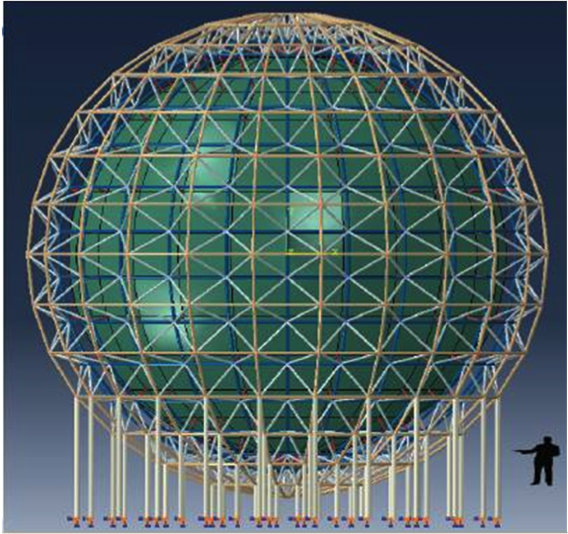}
    \figcaption{\label{fig1} Sketch of the Acrylic option for center detector design. }
    \end{center}
    
    \subsection{Balloon option}
    In this option the container of LS is made of nylon or fluoride-rich material, with acrylic board and support structure to hold the balloon, as shown in Fig.~\ref{fig2}. The buffer material between the steel tank and nylon ball is linear alkylbenzene (LAB), which serves as the solvent of LS.
    A steel tank holding LS, buffer material and PMTs is placed in the water pool.
    
    \begin{center}
    \includegraphics[width=7cm]{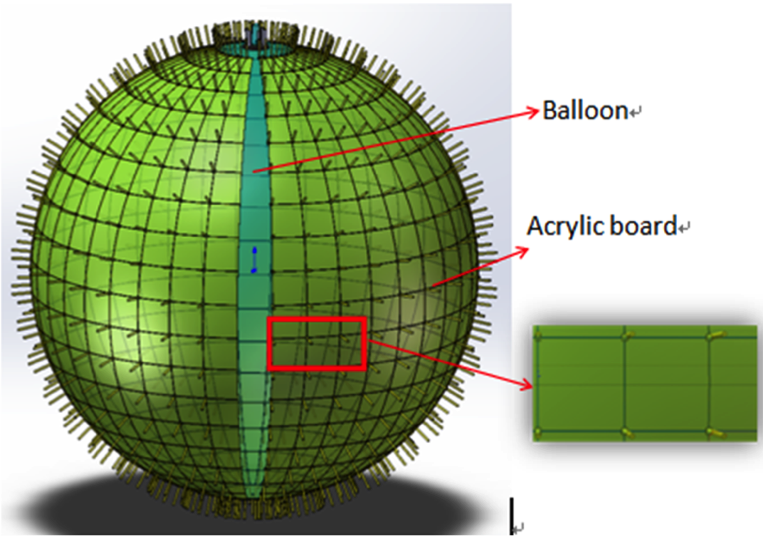}
    \figcaption{\label{fig2}   Sketch of the Balloon option for center detector design.}
    \end{center}
    
    \subsection{Module option} 
    The module option is proposed to reduce risk of big structures. The module is made of acrylic.
    Two kinds of encapsulations can be taken for different module sizes, either with only one PMT in each small module, or with a set of PMTs in each large module. 
    \subsubsection{Small module option}
    In this option, each module is equipped with a single PMT, as shown in Fig.~\ref{fig3}, are directly placed in the LS, with shielding material  filling up the module. The location of the modules is the same as the PMT arrangement in the acrylic option. 
   
    \begin{center}
    \includegraphics[width=8cm]{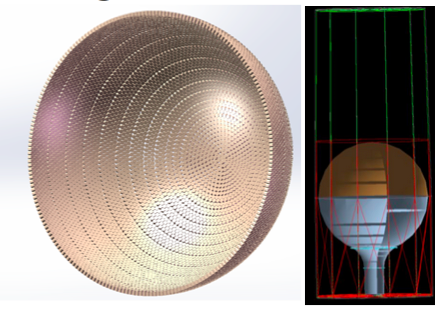}
    \figcaption{\label{fig3} Small modules and their arrangement in a half sphere.}
    \end{center}
       
    \subsubsection{Large module option}
    In the large module option, to fill up the whole sphere, the spherical surface is divided into many different triangles. 
    First, the surface of the sphere at radius of the PMT position is divided into 20 equal parts. Then each part is divided into triangles in ten different kinds of shapes. 
    The sizes of the modules are designed according to the size of the triangles. The numbers of PMTs are different in different  modules. 
    A double layer configuration of PMT has been considered to improve the coverage of PMT.
    
    \begin{center}
    \includegraphics[width=6cm]{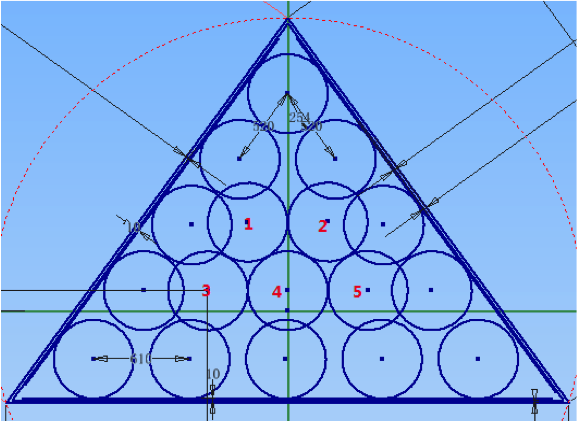}
    \figcaption{\label{fig4}  PMT arrangement in a big module}
    \end{center}

    As shown in Fig.~\ref{fig4}, the first layer of the module is filled with closely arrayed PMTs . The number of PMTs in the second layer is less than that in the first one. They are set to fill up the gaps between the PMTs in the first layer.
    The circles with a number represent the PMTs in the second layer.
   
    \begin{center}
    \includegraphics[width=6cm]{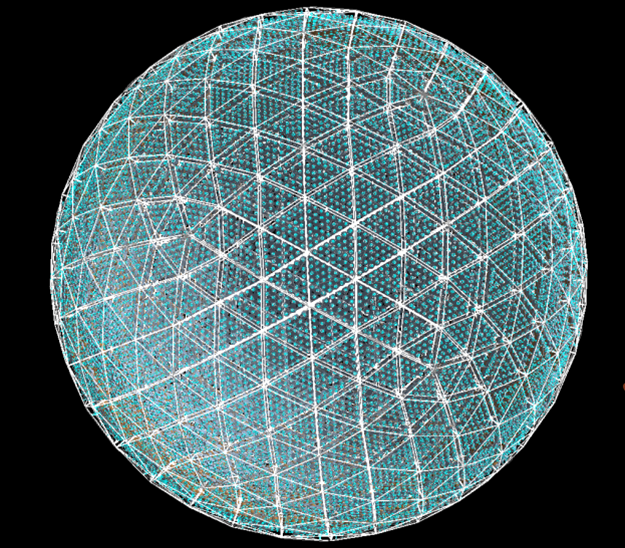}
    \figcaption{\label{fig5} The display of all PMTs in big module option.}
    \end{center}
   
   Fig.~\ref{fig5} shows the overall arrangement of the large module option using Geant4.   

   \section{Comparison of the simulation results of several options}
    
   The energy spectrum of the backgrounds due to natural radioactive elements overlaps with that of the antineutrino inverse beta decay(IBD). 
   When a combination two background events meet the selection criteria of IBD, they can form an accidental background and imitate a genuine neutrino event.

    Singles rate is an important factor to consider when selecting the detector option. 
    Small singles rate of the central detector is required.
    In the following, the singles means signals from radioactivity depositing $>$0.7 MeV of visible energy in LS.
    Some parameters are set to be consistent in these options, namely the location of the PMTs, the optical parameters of LS and the thickness of the buffer.
    We mainly compare the background from PMT glass in each option. The background from PMT glass is mainly due to the radioactive elements ($^{238}$U, $^{232}$Th, $^{40}$K).  
    If the schott glass\cite{lab9} is chosen, the radio-purities of glass are 22 ppb, 20 ppb and 3.54 ppb, respectively. 
    Fig.~\ref{fig6} shows the singles rates from PMTs in different options. 
    
    \begin{center}
    \includegraphics[width=8cm,height=6cm]{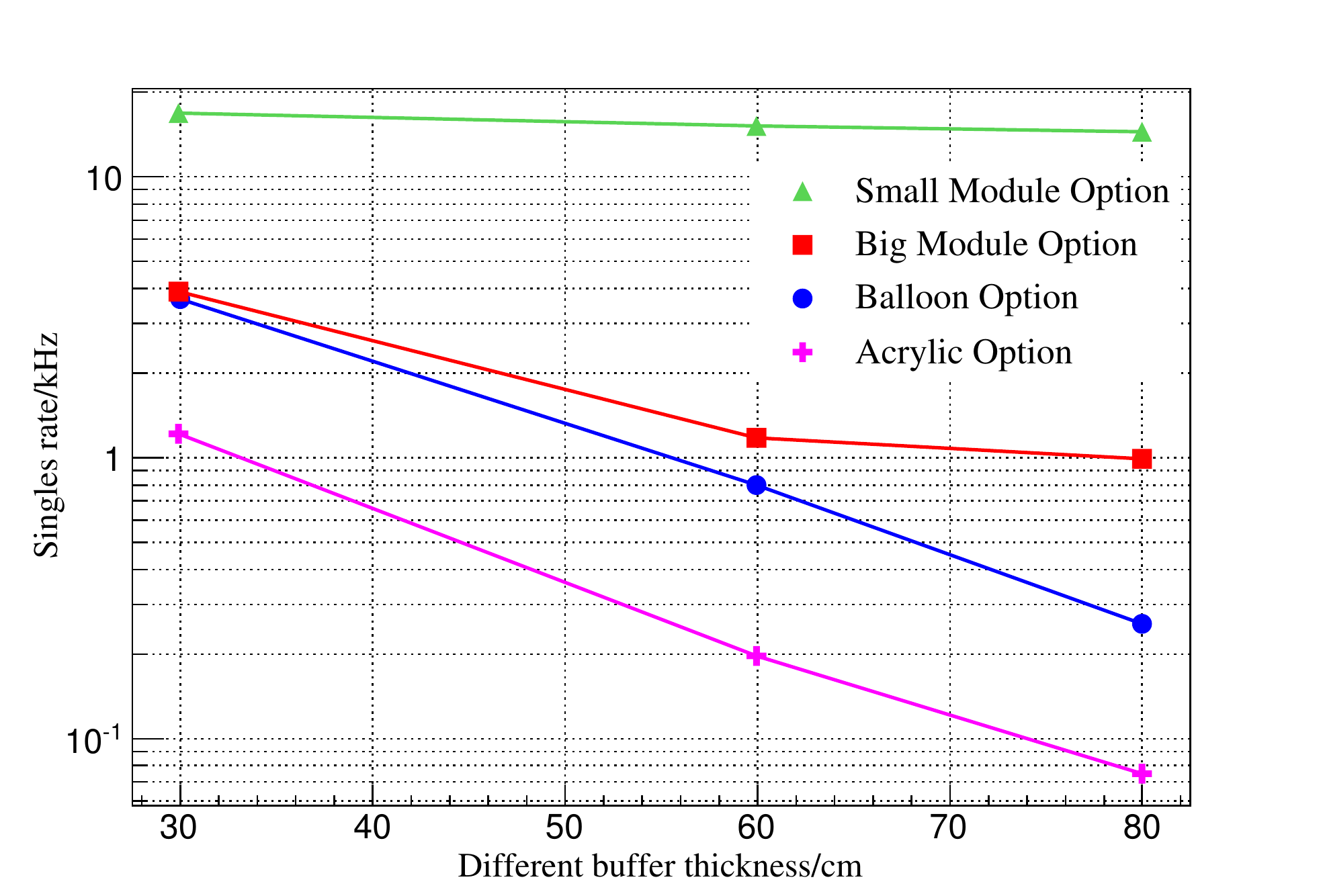}
    \figcaption{\label{fig6}  Simulation results of the radioactive background rates for different detector options.}
    \end{center}
    
    The singles rates of PMTs in acrylic and balloon option decrease exponentially with the increase of buffer thickness. However, the module options are not, which is because the gaps of each module are filled  with LS. Module option can't shield the radioactivity background from PMT glass as good as the other two options.
    Finally, the module option has been abandoned.
    Since the acrylic and balloon option are similar, in the following sections we just consider the acrylic option. 

    \section{Research on acrylic option}
    Natural radioactivity exists in the materials of JUNO detector components and its surroundings. Particular care needs to be taken to select low radioactive materials and to design the shielding to control radioactive backgrounds. For JUNO experiment, the radioactivity comes from various sources, such as $^{238}$U, $^{232}$Th,$^{40}$K. 
    In Table~\ref{tab1} and Table~\ref{tab3} the concentrations of primordial nuclides are given as mass fractions (g/g), using the conversion relation for the case of secular equilibrium:
      
    \begin{quote} 
      1 ppb $^{238}$U = 12.40 mBq/kg\\
      1 ppb $^{232}$Th = 4.05 mBq/kg\\
      1 ppb $^{40}$K = 271 mBq/kg\\
      \end{quote} 
    
    In simulation,  the GenDecay package is used as the radioactivity generators.
    It is assumed that the decay series is in secular equilibrium.
    The package use Evaluated Nuclear Structure Data File(ENSDF)\cite{lab10} to get the decay particles, half-life and branching ratios. 
    
    \subsection{The radioactivity of inner materials}
    The materials of the central detector in the acrylic option mainly include LS, acrylic, oxygen-free copper, stainless steel, as well as the glass of PMT. Based on the experience\cite{lab11,lab12,lab13} from the existing neutrino experiments, the specification on radioactivity of detector materials are listed in Table~\ref{tab1}.
    The buffer thickness in front of PMT determines the singles rate from PMTs, so it needs to be optimized by varying the thickness of water in our simulation.
    The singles rate of PMT is nearly 3 Hz when the water thickness is about 1.426 m. LS radius is 17.7 m. Acrylic thickness is 12 cm.
    The radius of the sphere where the center of PMTs is located is 19.5 m.
    
      \end{multicols}
      \begin{center}
      \tabcaption{ \label{tab1}  The proposed concentration of radioactive impurity in different detector materials.}
      \footnotesize
      \begin{tabular*}{170mm}{@{\extracolsep{\fill}}cccccccc}
      \toprule \hphantom{0} & $^{238}$U & $^{232}$Th & $^{40}$K  & $^{210}$Pb & $^{85}$Kr & $^{39}$Ar & $^{60}$Co\\
        \hline
          LS & $10^{-6}$ ppb & $10^{-6}$ ppb & $10^{-7}$ ppb & $1.4\cdot10^{-13}$ ppb & 50 $\mu Bq/m^3$ & 50 $\mu Bq/m^3$ & \~{} \\
          Glass & $22$ ppb & $20$ ppb & $3.54$ ppb & \~{} & \~{} & \~{} & \~{}  \\
          Acrylic & $10$ ppt  & $10$ ppt & $10$ ppt & \~{} & \~{} & \~{} & \~{}  \\
          Steel & $ 1.2$ mBq/kg & $8.0$ mBq/kg  & $13.4$ mBq/kg & \~{} & \~{} & \~{}& $2.0 $ mBq/kg \\
          Copper & $1.23$ mBq/kg  & $0.405$ mBq/kg & $0.0377$ mBq/kg & \~{} & \~{} & \~{} & \~{} \\
          \bottomrule
          \end{tabular*}
          \end{center}

          \begin{multicols}{2}
     Above external radioactivity can be rejected by proper fiducial volume cut since their energy deposits are mainly at the LS edge. Thus, the internal LS radio-purity is very important for the JUNO experiment and should be well controlled. 
     The fractional distillation process at the last step of raw LAB production and water extraction of the fluor are necessary to improve the radio-purity of raw LS materials. 
     There will be nitrogen protection during LS production, in order to suppress Radon contamination. 
     However, the residual Radon contamination will lead to non-equilibrium isotope $^{210}$Pb (and the subsequent$^{210}$Bi decay) which has a half life of 22 years. 
     In the JUNO experiment, the initial purity level of LS that can be achieved without distillation is shown in Table~\ref{tab1}.  
     After setting up the on-line distillation, we believe better purity level with improvement of two orders of magnitude can be achieved: $10^{-17}$ g/g for $^{238}$U/$^{232}$Th, $10^{-18}$ g/g for $^{40}$K and $10^{-24}$ g/g for $^{210}$Pb.
     In this work, the purity level of LS is set to the value without distillation. 

     Full Monte Carlo simulation is performed to obtain the singles rates from LS and other detector
     construction materials. The singles rates with different fiducial volume cuts are listed in Table~\ref{tab2}. A fiducial volume cut is necessary to reject the external radioactivity and thus to reduce the accidental background.
    
     \end{multicols}
     \begin{center}
     \tabcaption{ \label{tab2}  The inner singles rates (E$>$0.7 MeV) in different fiducial volume.}
     \footnotesize
     \begin{tabular*}{170mm}{@{\extracolsep{\fill}} c c c c c c c}
     \toprule Fiducial Cut(m) &LS(Hz) & Glass(Hz) & Acrylic(Hz)  & Steel(Hz) & Copper(Hz) &Sum(Hz) \\
       \hline
         R$<$17.7  &2.39 &2.43  &69.23  &0.89  &0.82  &75.76 \\
         R$<$17.6  &2.35 &1.91  &41.27  &0.66  &0.55  &46.74 \\
         R$<$17.5  &2.31 &1.03  &21.82  &0.28  &0.32  &25.76 \\
         R$<$17.4  &2.27 &0.75  &12.23  &0.22  &0.19  &15.66 \\
         R$<$17.3  &2.24 &0.39  &6.47   &0.13  &0.12  &9.35  \\
         R$<$17.2  &2.20 &0.33  &3.61   &0.083 &0.087 &6.31  \\
         R$<$17.1  &2.16 &0.23  &1.96   &0.060 &0.060 &4.47  \\
         R$<$17.0  &2.12 &0.15  &0.97   &0.009 &0.031 &3.28  \\
         \bottomrule
         \end{tabular*}
         \end{center}
         \begin{multicols}{2}
         
         \subsection{Radioactivity from water and rock}
         The natural radioactivity in detector surroundings mainly comes from water and rock. 
            
             \begin{center}
             \tabcaption{ \label{tab3}  The proposed concentration of radioactive impurity in water and rock.}
             \footnotesize
             \begin{tabular*}{70mm}{@{\extracolsep{\fill}}|c|ccc|}
             \hline
             Water &\multicolumn{3}{c|}{Rock}\\
               \hline
               $^{222}$Rn & $^{238}$U &$^{232}$Th &$^{40}$K \\
                 $0.2$ $Bq/m^{3}$ & 10 ppm & 30 ppm &5 ppm  \\
                 \hline
                 \end{tabular*}
                 \end{center}

                 In this work, we have only considered the $^{222}$Rn in water. The proposed concentration of radioactive impurity in water and rock are shown in Table~\ref{tab3}.
        
        \subsubsection{$^{222}$Rn in water}
         In a underground laboratory, radon concentration will reach an equilibrium when the decay balances with the emanation from the rock. 
         Radon is soluble in water. The equilibrium state for a radon-air-water system with respect to the diffusion process is described by the Ostwald coefficient the radon concentration in water is nearly 25\% of that in air at room temperature. 
         The circulation of water could bring radon to detector. When the radon concentration in air is $50$ $Bq/m^{3}$, its concentration in water is assumed to be equally $12.5$ $Bq/m^{3}$.
         The singles rate of $^{222}$Rn in water is so large that some methods should be taken to reduce the concentration of $^{222}$Rn in water.
        If the concentration can be limited to $0.2$ $Bq/m^{3}$, with good $N_{2}$ seal and sufficient anti-Rn liner on the water pool walls, then the singles rate is nearly 1.3 Hz when fiducial volume is 17.2 m.

         \subsubsection{The background in rock}
         Assuming that the radioactivity of rock at JUNO experimental site is similar to that measured
         at Daya Bay site: ∼10 ppm for $^{238}$U, ∼30 ppm for $^{223}$Th and ∼5 ppm for $^{40}$K. Since a full Monte
         Carlo simulation would be extremely time consuming, a numerical calculation is performed
         to estimate the effect of the rock radioactivity.

         In this method, 50 cm rock around the water pool is divided into many small parts.
         After finishing dividing, the singles rate of each small part is calculated separately.
         There are five steps involved in each calculation. 
         
         \begin{enumerate}         
         \item Calculate each small part's mass to obtain the initial event rate. 
         \item Calculate the total event rate of $\gamma$ (betas are not counted because they can't travel through water). The ratio of $\gamma$ is gotten from simulation. Take $^{232}$Th, for instance. There are 665797 $\gamma$ from $10^{6}$ decays, as illustrated in Fig.~\ref{fig7}.
         \item Calculate the event rate of $\gamma$ at each energy point. The ratio of $\gamma$ versus energy also obtained from simulation. In the case of $^{232}$Th, they are shown in Fig.~\ref{fig7}.
         
         \begin{center}
         \includegraphics[width=8cm,height=6cm]{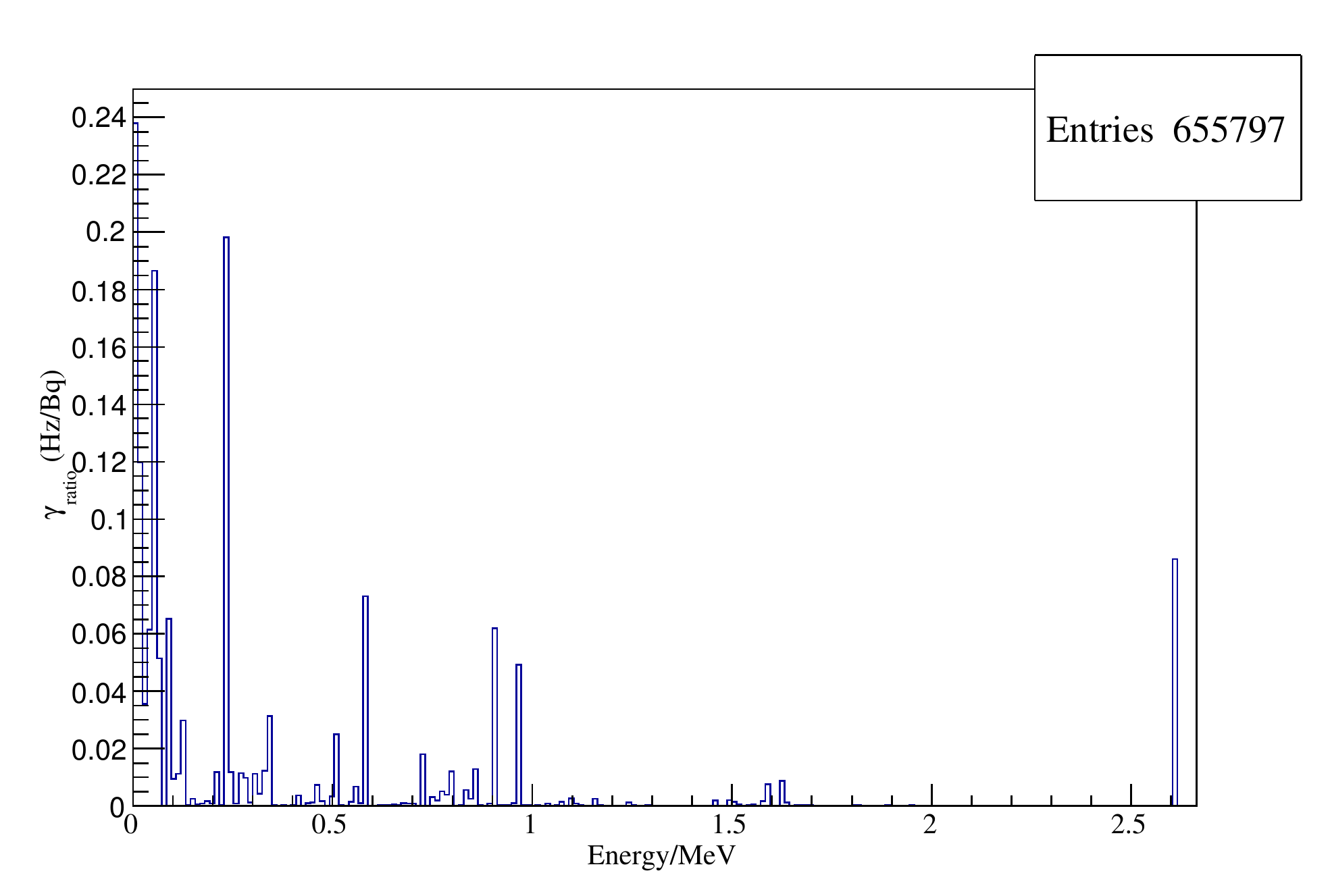}
         \figcaption{\label{fig7}  The $\gamma_{ratio}$ versus energy. $\gamma_{ratio}$ is the ratio of the $\gamma$ event rate in (Hz) with different energy of $\gamma$ to the decay rate of $^{232}$Th chain in (Bq). }
         \end{center}
         
         \item The effective solid angle($\Omega_{eff}$) of each voxel to the LS detector is calculated by taking into account the attenuation of different water thickness.
        
         \begin{equation}
         \label{one}
         \Omega_{eff} = \sum_{i}(\frac{\Delta s_{i}cos \alpha_{i}}{L_{i}^{2}} \cdot e^{-L_{i}\cdot A_{w}})/4\pi,
         \end{equation}
         As shown in Fig.~\ref{fig8}, the surface is divided into many small parts. $\Delta\beta$ is the corresponding angle of each small surface. $\alpha_{i}$ is the angle between the short dash line and long dash line. $\beta_{i}$ is the angle between the long dash line and the thick solid line.
         $L_{i}$ is the distance between rock and the detector at different angles.
         D is the minimum distance between rock and the central detector.
         $A_{w}$ is the attenuation coefficients of water in different energy and water length. 
        
         \begin{equation}
         \label{two}
         \Delta s_{i} = 2 \pi R^{2} \cdot sin\beta_{i} \cdot \Delta \beta,
         \end{equation}
         
         \begin{center}
         \includegraphics[width=5.5cm,height=5cm]{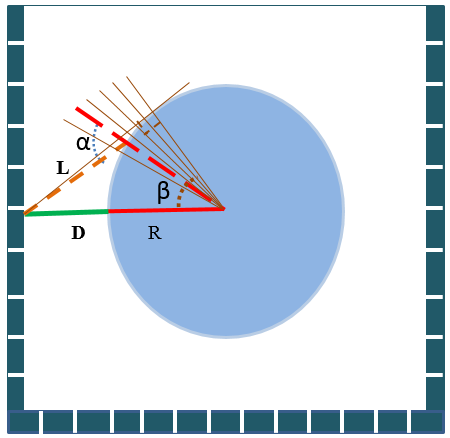}
         \figcaption{\label{fig8}  The schematic of these variables in estimation. }
         \end{center}
        
        \item We can calculate the event rates of each small part after attenuation.
         \end{enumerate}
         
         \begin{equation}
         \label{three}
         R_{parts} = \sum_{E} R_{E} \cdot \Omega_{eff} \cdot R(E),
         \end{equation}
         
         \begin{equation}
         \label{four}
         R(E) = e^{-L_{r}A_{r}},
         \end{equation}

         $R(E)$ is the attenuation in rock.
         Where $A_{r}$ is the attenuation coefficients of rock in different energy, $L_{r}$ is the distance of $\gamma$ penetrates rock.

         In this empirical calculation, the attenuation coefficients are very important.
         In order to get the attenuation coefficients, a simple model has to be simulated.
         In this model, each different energy $\gamma$ penetrates materials vertically with different thickness. 
         The initial number of $\gamma$ is $N_{init}$, and the thickness of material is L.
         $N_{rec}$ is the number of the $\gamma$ with energy greater than 0.7 MeV after passing through the materials.
         Then, we can calculate the attenuation coefficients with the formula as below.
         
         \begin{equation}
         \label{six}
         A_{w}=-ln(N_{rec}/N_{init})/L,
         \end{equation}

         As shown in Fig.~\ref{fig9}, we can get the coefficients of each different energy in the case of different thicknesses of water. 
        
         \begin{center}
         \includegraphics[width=8cm,height=6cm]{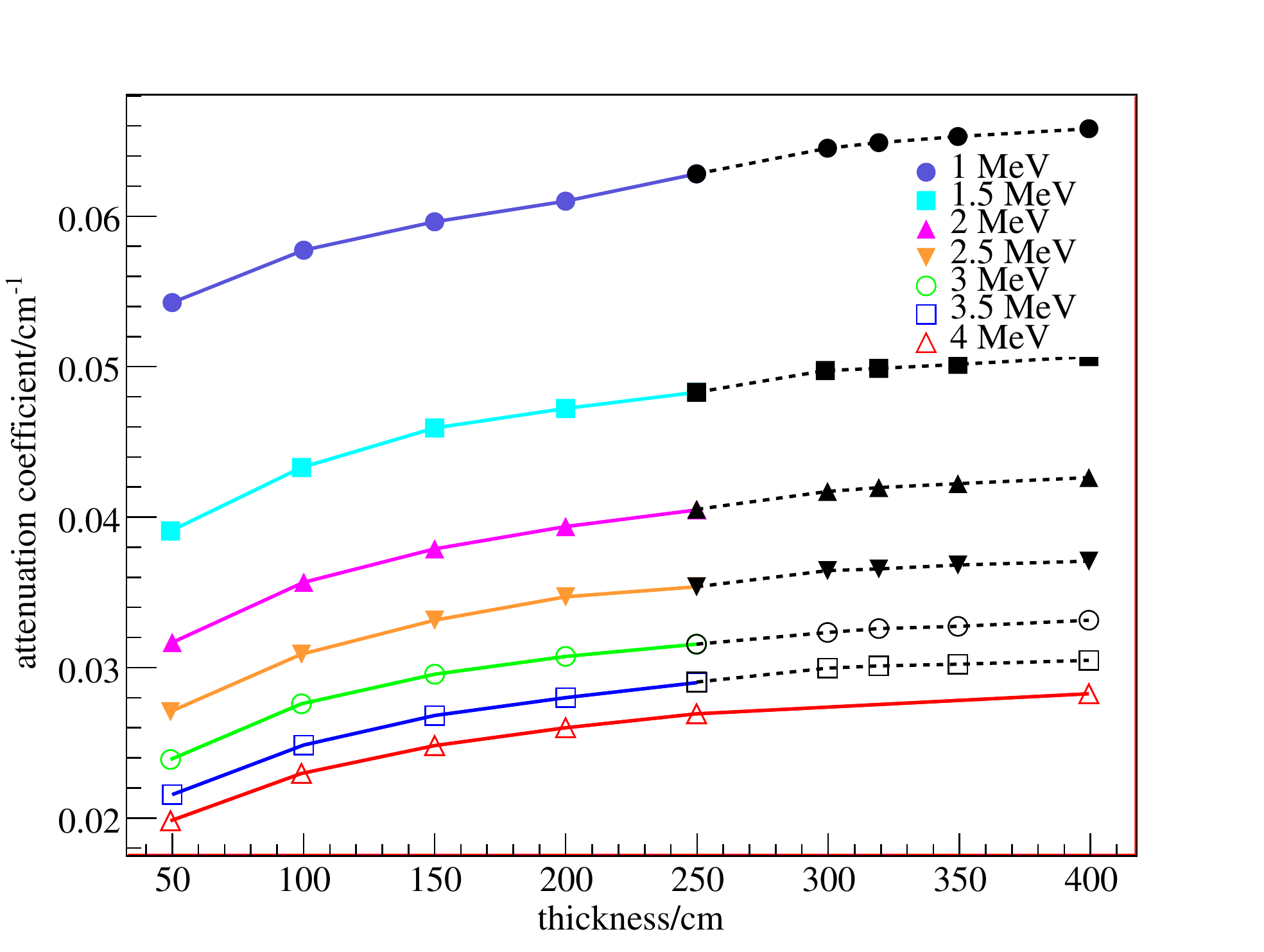}
         \figcaption{\label{fig9} Attenuation coefficients of $\gamma$ in Eq.(1) with different energy in the case of different thicknesses of water. The different marker styles represent different energies. The dashed lines are estimated by the extrapolation method, assuming that the coefficient ratios between different energies are fixed.}
         \end{center}

         We use extrapolation method to calculate the attenuation coefficients when the thickness is greater than 2.5 m, because it is difficult to get these coefficients in simulation. 
         With these attenuation coefficients we can calculate the final signals rates in different water thicknesses.
         A real simulation has been performed to verify the validity of the estimation method. 
         In this simulation, the rock thickness is set to be the same as that used in the estimation, and the water thicknesses are 2 m and 2.5 m. 
       
       \begin{center}
       \tabcaption{ \label{tab4}  Comparison of the results between simulation and estimation with different water thickness.  }
       \footnotesize
      \begin{tabular*}{80mm}{@{\extracolsep{\fill}}|c|c|c|c|}
         \hline
        & &Simulation(Hz) & Estimation(Hz)\\
           \cline{1-4}
           &$^{40}$K
           &161.2 
           &68.9 \\
             \cline{2-4}
             2 m water
             & $^{232}$Th
             & 981.0 
             & 829.5 \\
               \cline{2-4}
               & $^{238}$U
               & 162.2 
               & 154.1 \\
                 \cline{1-4}
                 & $^{40}$K
                 & 9.0 
                 & 4.9 \\
                   \cline{2-4}
                   2.5 m water
                   & $^{232}$Th
                   & 143.2 
                   & 133.8 \\
                     \cline{2-4}
                     &  $^{238}$U
                     &  35.6 
                     &  17.6\\
                       \hline
                       \end{tabular*}
                       \end{center}

                       Comparison of simulation and estimation is shown in Table~\ref{tab4}.
                       We can tell from the comparison that the estimation is
                       similar to the simulation results. Thus this method can be used to calculate singles rate of rock.
                       The singles rates of $^{40}$K/$^{232}$Th/$^{238}$U are estimated to be 0.0742 Hz, 6.739 Hz, 0.613 Hz, respectively, when the thickness of water is 3.2 m.
                       These results satisfied the requirements for the JUNO experiment. 
             
             \subsection{The accidental background from natural radioactivity}
                  
             The sum singles rates with different fiducial volume cut are displayed
             in Table~\ref{tab5}.
             Base on this table, the radius of fiducial volume is set to 17.2 m. The total singles rate is 8.6 Hz.

             \begin{center}
             \tabcaption{ \label{tab5} The sum singles rates(E$>$0.7 MeV) with
             different fiducial volume cut.}
             \footnotesize
             \begin{tabular*}{80mm}{c@{\extracolsep{\fill}}ccccc}
          \toprule    \tabincell{c}{Single Rate\\(Hz)} &\tabincell{c}{Detector\\ components} &Water &Rock &Sum  \\
               \hline
                R$<$17.7 m  &75.76 &15.94 &7.42 &99.12\\
                R$<$17.6 m  &46.74 &11.00 &4.47 &62.21\\
                R$<$17.5 m  &25.76 &6.58  &2.90 &35.24\\
                R$<$17.4 m  &15.66 &3.84  &2.02 &21.52\\
                R$<$17.3 m  &9.35 &2.20  &1.41 &12.96\\
                R$<$17.2 m  &6.31  &1.31  &0.98 &8.60\\
                R$<$17.1 m  &4.47  &0.78  &0.68 &5.93\\
                R$<$17.0 m  &3.28  &0.46  &0.42 &4.16\\
               \bottomrule
                 \end{tabular*}
                 \end{center}

                 Base on simulation of natural radioactivity background from different detector materials, we can get the scatterplot of their deposited energy of radioactivity background versus LS radius, as shown in Fig.~\ref{fig10}. 
                 With this scatterplot we can calculate the number of
                 accidental background from them in fiducial volume.

                     \begin{center}
                     \includegraphics[width=8cm,height=6cm]{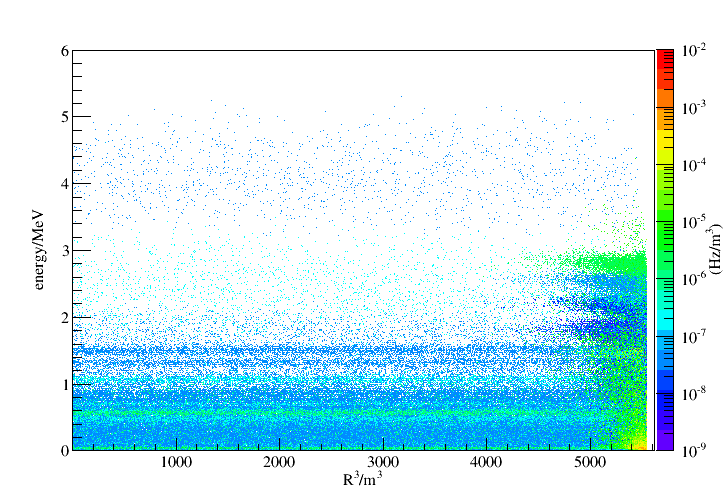}
                     \figcaption{\label{fig10}  Scatterplot of deposited energy versus LS radius.}
                     \end{center}
                     
                     According to the singles rate, a random sample of some background events were generated based on the scatterplot. Then, certain events meeting the criteria listed below were selected. 
                    
                     \begin{enumerate} 
                     \item the prompt energy cut 0.7 MeV $<$$E_{p}$ $<$ 12 MeV;
                     \item the delayed energy cut 1.9 MeV $<$ $E_{d}$ $<$ 2.5 MeV; 
                     \item time difference between the prompt and delayed signal: 1.0 $\mu s$ $<$ $\Delta T$ $<$1.0 ms;
                     \item the prompt-delayed distance cut $R_{p-d}$ $<$1.5 m;  
                     \end{enumerate} 
                     
                      Finally, we can get that the accidental background rate in fiducial volume is nearly 1.1/day. This is acceptable compared with the expected signal rate of 60/day IBD.
                     
               \section{Conclusion}      

                     JUNO experiment is designed to operate with very low
                     radioactivity backgrounds. From the Geant4 simulation
                     results, we know that with the buffer thickness between
                     PMT and LS is 1.5 m, buffer thickness between rock and LS
                     is 3.2 m, and also with the requirements for radio-purity
                     of materials in the JUNO detector which shown in
                     Table~\ref{tab1} and Table~\ref{tab3}, cut LS fiducial
                     volume radius from 17.7 m to 17.2 m, the accidental
                     background induced by natural radioactivity in JUNO
                     central detector is 1.1/day, which is 1.8\% of the
                     expected IBD signal events(60/day). The results of the present study will provide an important basis for optimization of the JUNO design.

                     \acknowledgments{The authors gratefully thank members of
                     JUNO central detector group for their valuable
                     discussions, and sincerely thank Xiaoyan Ma, Xiaohui Qian, and Jiajun Hao for their help on PMT arrangement.}

                     \end{multicols}
                     \vspace{-1mm}
                     \centerline{\rule{80mm}{0.1pt}}
                     \vspace{2mm}

                     \begin{multicols}{2}
                     
                     \end{multicols}

                     \end{document}